\documentclass[aps,pra,reprint,showpacs,superscriptaddress]{revtex4-1}

\usepackage{graphicx}
\usepackage[utf8]{inputenc}
\usepackage{caption}
\usepackage{siunitx}
\usepackage{amsmath}
\usepackage{gensymb}
\usepackage{amssymb}
\usepackage{enumerate}
\usepackage{color}
\usepackage{hyperref}
\usepackage[american]{babel}
\usepackage{subcaption}
\usepackage{etoolbox}
\usepackage{floatrow}
\usepackage{nicefrac}

\patchcmd{\thebibliography}{\section*{\refname}}{}{}{}

\newcommand{\Hbar}{$\bar{\hbox{H}}$}
\newcommand{\Dbar}{$\bar{\hbox{D}}$}
\newcommand{\pbar}{$\bar{\hbox{p}}$}
\newcommand{\dbar}{$\bar{\hbox{d}}$}
\newcommand{\nbar}{$\bar{\hbox{n}}$}
\newcommand{\pos}{$\bar{\hbox{e}}^+$}

\DeclareSIUnit\inch{in.}
\DeclareSIUnit\division{div}

\begin{document}

\preprint{APS/123-QED}

\title{Perspectives from a cold antideuteron beam in the AD/ELENA facility}

\author{Ruggero Caravita}

\date{\today}

\begin{abstract}
This article reviews the perspectives opened by the development of a low-energy antideuteron (\dbar{}) beam: precision measurements of \dbar{} properties; formation and spectroscopic analysis of antideuteronic atoms and antideuterium; the pioneering synthesis of heavier anti-elements in particle traps and at low energies. Some practical aspects of generating a low-energy \dbar{} beam in the existing AD/ELENA facility are also discussed.
\end{abstract}

\maketitle{}

\section{Introduction}

Recent years advances in antimatter physics would have not been possible without the development and the refinement of the techniques for producing, storing, and cooling antiprotons (\pbar{}), at the heart of the two decelerators presently constituting the Antimatter Factory at CERN: the AD and the recently added ELENA. The AD/ELENA facility, with its current availability of about ten million \pbar{}s per \SI{2}{\minute} at a \SI{100}{\kilo\electronvolt} energy, has been enabling the accurate study of the properties of the \pbar{} (charge, mass, magnetic moment) and \Hbar{} (charge, energy levels, and gravitational coupling) and to conduct accurate tests of CPT symmetry and the Weak Equivalence Principle.

On the other hand, as of today, experiments employing cold anti-nuclei are still out of reach. This is due to the technical difficulties in both directly producing anti-nuclei (except in minuscule quantities in very high-energy collisions as in \cite{ALICE2022Antinuclei}) or proceed through antineutron (\nbar{}) capture on targets of cold antiprotons. Indeed, low-energy antineutron beams are unavailable as well, due to the absence of any electric charge to manipulate them after production. The only anti-nucleus which can be produced in modest amounts is the antideuteron (\dbar{}, first observed from proton collisions in fixed-target experiments in 1965 \cite{Zichichi1965, Dorfan1965}). Several production mechanisms have been discussed with a variety of efficiencies (from $0.1$ to $10^{-5}$) and momentum/energy distributions (see \cite{Moehl1982, Koch1984, Johnson1989, Koch1989}). Being charged, the \dbar{} can be stored and cooled in similar ways as the antiproton. 

In this short review, we reassess the prospects offered by the development of a cold antideuteron beam. In Section \ref{section_antideuterons}, we review the different measurement which can be performed on the antideuteron itself. In Section \ref{section_antideuteronicatoms}, we review the physics cases offered by the formation and study of antideuteronic atoms. In Section \ref{section_antideuterium}, we review antideuterium formation and measurements. In Section \ref{section_heavierantinuclei}, we review the perspectives in producing heavier anti-elements in trap through fusion processes enabled by antideuterons. Finally, in Section \ref{section_dbarinad}, we briefly discuss the feasibility of realizing a low energy \dbar{} beam in the current AD/ELENA facility, without the pretense that it is an accurate feasibility study, which should be the subject of a more specific work. 

\section{Antideuterons}
\label{section_antideuterons}



\subsection{Precision antideuteron mass measurement}
\label{subsect_dbarmass}

The antideuteron mass has been determined by direct measurements in fixed target experiments at the percent level ($ 1867 \pm 80 \, \si{\mega\electronvolt}/c^2 $) \cite{Zichichi1965} \footnote{no explicit reference to measurements done at colliders found yet.}. The deuteron mass, in contrast, is know by precision measurements of its charge-to-mass ratio in Penning traps (i.e., a direct mass measurement assuming electric charge quantization), with 8 ppt uncertainty \cite{Rau2020}. 

Similar charge-to-mass ratio precision measurements with charged antimatter systems at ppt level of precision are possible as well, as recently demonstrated by the BASE collaboration reaching 16 ppt in the determination of the \pbar{} charge-to-mass-ratio \cite{BASE2022}. The same approach could be used to perform high precision mass measurements with single \dbar{} in Penning traps, requiring only a handful antideuterons: typically, 30 \pbar{}s are sufficient for a yearly run of BASE \cite{BASEPrivateConv}. Such a measurement would constitute as well a stringent test of CPT and Lorentz invariance by direct comparison with deuteron results. Assuming \dbar{} at \SI{100}{\kilo\electronvolt} kinetic energy and a $ 50 \% $ trapping efficiency, it is reasonable to estimate that these measurements will become possible with integrated $ \propto 100 $ delivered \dbar{} in a run lasting two weeks ($ \sim 10000 $ AD shots), corresponding to 0.01 \dbar{} $ \mathrm{shot^{-1}} $.

\subsection{Magnetic moment of the antideuteron}

The antideuteron magnetic dipole moment has not been experimentally determined yet. On the contrary, the deuteron magnetic dipole and electric quadrupole moments are known at the level of 0.5 ppb and 700 ppm respectively \cite{Krane1987}, whereas experiments to determine the electric dipole moment of deuterons have been only recently proposed \cite{Anastassopoulos2008}. 

Similar flux considerations apply also to the measurement of the \dbar{} magnetic moment of as these regarding the \pbar{} charge-to-mass ratio precision measurements. Indeed, measurements of the \pbar{} magnetic moment have reached a precision of ppb \cite{Smorra2017} by using a similar of trapped antiprotons in an entire experimental run as for precision mass measurements. Thus,  it is reasonable to estimate that these measurements will become possible with integrated $ \propto 100 $ delivered \dbar{} in a run lasting two weeks ($ \sim 10000 $ AD shots), corresponding to 0.01 \dbar{} $ \mathrm{shot^{-1}} $.

\subsection{Measurement of the antideuteron binding energy}

The antideuteron binding energy has not been measured yet. On the contrary, the deuteron binding energy has been directly measured at the 1 ppm level by spectroscopy of the $\gamma$ resulting from coalescence in neutron-proton collisions \cite{Krane1987}.

A measurement of the \dbar{} binding energy cannot be realized by the same method due to the unavailability of a cold \nbar{} beam. However, alternative viable methods borrowed from the determination of the deuteron binding energy are the mass subtraction method and the photodissociation by $\gamma$ rays method \cite{Krane1987}.

The mass subtraction method relies on the precise knowledge of the \dbar{} mass, the \pbar{} mass and the \nbar{} mass to subtract them. The \pbar{} mass is known at 10 ppt level \cite{BASE2022}; the \nbar{} mass is known at the 50 ppm level \cite{Bressani2003}; a way to determine the \dbar{} mass with high precision is discussed in \ref{subsect_dbarmass}. Assuming the final accuracy of this method to be limited by the \nbar{} mass ($939.485 \pm 0.051 \, \si{\mega\electronvolt}$), the projected maximum sensitivity of this method is around $2 \%$ with just a handful of \dbar{} (i.e., with an estimated initial flux of 0.01 \dbar{} $ \mathrm{shot^{-1}} $).

The photodissociation method is built on top of the $ \bar{d}(\gamma,\bar{n})\bar{p} $ process where an energy tunable,  $ \approx \SI{2.2}{\mega\electronvolt}$ $ \gamma $ is used to ionize the antideuteron at threshold. Accuracies of 0.1 \% have been reported in literature for the deuteron using the matter-counterpart process $d(\gamma, n)p$ \cite{Krane1987}; recent experiments are also ongoing at the ELBE facility using bremsstrahlung radiation from an electron LINAC \cite{Hannaske2016}. 

The mass subtraction method is more keen to provide precision results of the \dbar{} binding energy, provided that a path forward to measure more accurately the \nbar{} mass is envisaged. This may come from the synthesis of a lower-energy and more monochromatic \nbar{} beam as in the past (see \ref{subsect_nbarbeam}).





\section{Antideuteronic atoms}
\label{section_antideuteronicatoms}

\subsection{Formation of antideuteronic atoms}

Antideuteronic atoms, i.e. bound states of an antideuteron an a positive ion, have been discussed in literature \cite{Koch1984} although not being reported so far.  Antideuteronic atoms can be formed at low energy either by letting \dbar{} in a trap interact with a low-pressure buffer gas, or by co-trapping \dbar{} with anions subsequently photodetached by a laser. The antideuteron is expected to strip off electrons from the atom it is attached to $\sqrt{2}$ times more efficiently than \pbar{}s \cite{Koch1989} while falling towards the nucleus and producing an X-ray cascade. In the case of $Z < 30$ nuclei, the antideuteron will ultimately get absorbed by the nucleus and annihilate, offering a testing ground for antideuteron-nucleus interactions \cite{Koch1989}. For $Z > 30$ nuclei, the antideuteron will undergo Coulomb dissociation into a \pbar{} and a \nbar{} by the strong nuclear field \cite{Baur1972, Ericson1975, Koch1989}. The Coulomb dissociation of a bound \dbar{} would be observable for the first time \cite{Koch1984}. The resulting antiproton bound to the nucleus will then proceed stripping it off, as in a regular antiprotonic atom. In both cases, nuclear fragments resulting from either the \dbar{} or the \pbar{} annihilation in the nucleus can be trapped in a nested trap configuration and identified by means of time-of-flight spectroscopy. 

A special case of antideuteronic atom is antideuteronium, the equivalent of protonium (the two-body bound state of an antiproton and a proton) made of an antideuteron and a proton. It is an effective two-body system (the antideuteron cannot break up in this very low Z \cite{Ericson1975}) and its energy levels can be calculated accurately. The advantage with other antideuteronic atoms is that, if synthesized directly from protons, the X-ray cascade of antideuteronium from its Rydberg levels (like that of Pn) will be hydrogen-like i.e., free of effects linked to inner filled atomic shells. 

Experiments with antideuteronic atoms require small amounts of \dbar{}. In principle, all \dbar{} interacting with either a buffer gas or laser photodissociated ions will end up forming antideuteronic atoms before annihilation. Thus, necessary flux is only bound to the detection system employed. As a reference, experiments performed by the AEgIS collaboration in forming antiprotonic atoms produced a statistically significant evidence of \pbar{}-gas interactions with about 5000 \pbar{} \footnote{unpublished}, setting the flux necessary for this experimentation to around 1 \dbar{}/shot for the necessary statistics to be collectable in two weeks of measurements with 50 \% trapping efficiency.  

\subsection{X-ray spectroscopy of antideuteronic atoms}

Once antideuteronic atom formation is established by one of the mechanisms discussed above, the spectrum of the X-rays emitted during the cascade of the \dbar{} towards the nucleus can be observed by X-ray spectroscopy, providing accurate insights on the energy levels. 

A shift in the X-ray lines of the cascade in the order of part-per-mil due to the antideuteron polarization in the cascade is also expected for $Z > 30$ nuclei \cite{Ericson1975}. The effect is in fact most pronounced towards the end of the cascade, when the \dbar{} approaches the dissociation threshold. Here the emitted X-ray energies are in the order of tens of \si{\kilo\electronvolt}. For example, for $ Z = 90 $, break-up of the \dbar{} is expected at $ n = 18 $ (as in \cite{Ericson1975}) and the energy of the hardest X-ray (that of the $ 19 \rightarrow 18 $ transition) is of \SI{14.2}{\kilo\electronvolt}. An experiment aiming at observing such shift should be able to measure the center of the line with per-mil accuracy, and have an energy resolution of $ \ll \SI{1.0}{\kilo\electronvolt} $ to resolve the individual lines. 

This kind of experiments will require higher amounts of \dbar{} to construct X-ray spectra with sufficient statistics. Taking as a reference X-ray spectra of Protonium spectroscopy at LEAR \cite{Welsh1989}, order of $ 10^5 $ produced X-rays are necessary (high detection efficiencies were obtained with X-ray drift chambers, see \cite{Gastaldi1978}). This corresponds to around 10 \dbar{}/shot for the necessary statistics to be collectable in two weeks of measurement time. 

\subsection{Low energy antineutron beams from high-Z antideuteronic atoms}
\label{subsect_nbarbeam}

Antideuteronic atoms with $ Z > 30 $ (i.e., in the regime in which the \dbar{} will undergo Coulomb dissociation \cite{Baur1972}) offer an interesting opportunity to produce $ \propto \si{\mega\electronvolt} $ \nbar{} beams. Indeed, for such heavy nuclei, the dissociation of the antideuteron will occur far from the nucleus, preventing strong annihilation effects. The resulting antiproton from the dissociation is bound to the nucleus by Coulomb attraction continuing the X-ray cascade, whereas the antineutron is freely emitted. The energy of the produced \nbar{} can be worked out following \cite{Baur1976}. It is given by the conservation of energy 

\begin{equation}
E_{\bar{n}} = - |\epsilon_b| - |E_{\bar{d}}| + |E_{\bar{p}}| 
\end{equation}

where $ \epsilon_b = \SI{2.22}{\mega\electronvolt} $ is the \dbar{} binding energy of the initial \dbar{}, $ E_{\bar{d}} $ is the hydrogenic energy of the initial nucleus bound to the \dbar{}, $ E_{\bar{p}} $ is the hydrogenic energy of the final nucleus bound to the \pbar{}

\begin{equation}
\begin{aligned}
 E_{\bar{d}} &= -\SI{13.6}{\electronvolt} \, \frac{m_{\bar{d}}}{m_e} \frac{Z^2}{n_{\bar{d}}^2} = \SI{50}{\kilo\electronvolt} \frac{Z^2}{n_{\bar{d}}^2} \\
 E_{\bar{p}} &= -\SI{13.6}{\electronvolt} \, \frac{m_{\bar{p}}}{m_e} \frac{Z^2}{n_{\bar{p}}^2} = \SI{25}{\kilo\electronvolt} \frac{Z^2}{n_{\bar{p}}^2}
\end{aligned}
\end{equation} \, .

Here $ m_{\bar{d}} $ and $ m_{\bar{p}} $ are the \dbar{} and \pbar{} masses respectively and $ n_{\bar{d}} $ and $ n_{\bar{p}} $ the principal quantum numbers associated to the initial and final bound states, respectively. For $ Z = 90 $, the \dbar{} breakdown threshold occurs at $ n^{(90)}_{\bar{p}} = 6 $ and $ n^{(90)}_{\bar{d}} = 13 \div 16 $, corresponding to \nbar{} emitted with $ E^{(90)}_{\bar{n}} = 1.0 \div 1.8\, \si{\mega\electronvolt} $ energy. For $ Z = 30 $, the \dbar{} breakdown threshold occurs at $ n^{(30)}_{\bar{p}} = 2 $ and $ n^{(30)}_{\bar{d}} = 7 \div 9 $, corresponding to \nbar{} emitted with $ E^{(30)}_{\bar{n}} = 2.5 \div 2.8 \, \si{\mega\electronvolt} $ energy. Another interesting aspect of this method is its efficiency: for $ Z \approx 90 $, close to all \dbar{} end up tore into \pbar{} and \nbar{} with the \nbar{} emitted. Polarization effects of the initial antideuteronic atom may also influence the angular distribution of emitted \nbar{} due to conservation of angular momentum: a topic deserving a more dedicated study. An experimental setup leveraging on this process could be realized by loading high-Z anions in a Penning trap from a Middleton-type Cs sputtering source \cite{Middleton1983}. Co-trapping them with \dbar{} and neutralizing the anion by on-threshold laser photo-detachment will create clear conditions for a neutral atom to capture \dbar{} and initiate the cascade. The \nbar{} yield is bound to the number of available \dbar{}, as the number of available anions will exceed it by several orders of magnitude.

\section{Antideuterium}
\label{section_antideuterium}

\subsection{Antideuterium production}

Antideuterium (\Dbar{}) formation is expected to work the same way as antihydrogen (\Hbar{}) production, apart from slightly different reaction rates \footnote{yet to be calculated}, consequence of their mass difference. In order to form antideuterium, one can proceed by three-body recombination (3BR) with a positron plasma in a nested trap as developed for \Hbar{} in \cite{Amoretti2002}. Plasmas of $ \approx 10000 $ \dbar{} would be necessary to reach the same signal-to-noise ratio as in \cite{Amoretti2002}, although a production signal may be observable already with a lower statistics. Nevertheless, assuming 6 hours stacking from AD/ELENA (usually among the longest stacking runs that can be performed without incurring into variations of the beam conditions), this flux corresponds to around 100 \dbar / shot delivered, with 50 \% trapping efficiency. 

\subsection{Antideuterium trapping and spectroscopy}

The successful construction and operation of a deuterium maser with absolute frequency uncertainty $ \approx \SI{1}{\milli\hertz} $ \cite{Wineland1972} implies that high-precision spectroscopy of the antideuterium hyperfine structure is a realistic possibility, once antideuterium can be formed and stored in a suitable trap. 

From the technical point of view, antideuterium trapping will present many similarities as antihydrogen. Assuming to trap \Dbar{} with the same technique the ALPHA collaboration has developed for antihydrogen \cite{Andresen2010}, one has to account that the trap depth is only of \SI{0.54}{\kelvin}, and just a small fraction of the produced \Hbar{} (\Dbar{}) will in fact be trappable. For \Hbar{}, the ratio between trappable \Hbar{} and initial \pbar{}s is about 1/10000: 10 \Hbar{} trapped out of $ 9 \cdot 10^4 $ initial \pbar{} in the mixing trap \cite{Ahmadi2017}. This corresponds to one order of magnitude more flux necessary to obtain the necessary \dbar{} numbers in a 6 hours stacking run compared to the case before, setting the threshold of feasibility to about 1000 \dbar / shot delivered.

\subsection{Gravity with antideuterium: constraining $ B - L $ interactions}

Unlike antihydrogen, antideuterium has a net $B - L$ charge, thus it is sensitive to hypothetical BSM forces coupling to it and it would experience a violation in the Weak Equivalence Principle if there would be any long-range $ B - L $ interactions with the Earth. 

A class of BSM theories which could produce such long-range forces distinguishing matter and antimatter are, for example, gauged $ B - L $ theories i.e., those where the $B-L$ symmetry of the Standard Model is incorporated into the full gauge group by an extra $ U(1)_{B-L} $ term, linked to a new gauge field mediated by a spin-1 boson Z' \cite{Charlton2020}. These forces fall into the standard parametrisation of the modified Newtonian gravitational potential derived from single-boson exchange:

\begin{equation}
V(r) = -\frac{G_\infty m_1 m_2}{r} \left( 1 - \alpha e^{-{r/\lambda}} \right)
\end{equation}

The exchange force here is attractive when the $ B - L $ charges of the interacting bodies are opposite, and it is repulsive when the charges are the same. Existing constraints on the strength of such a new interaction over the entire range of the Z' mass can be found in \cite{Heeck2014}. In the case of long-range forces (corresponding to light Z' $ m \ll meV $), existing EP tests on ordinary matter constrain the coupling to $ < 10^{-45} $. To what degree of generality such constraints do limit also potential $ B - L $ interactions happening on antimatter, it is yet to be understood in more detail.  

Nevertheless, a test of the Weak Equivalence Principle can be obtained with \Dbar{} similarly as how it was recently obtained with \Hbar{} \cite{Anderson2023}. The amount of necessary \Dbar{} atoms is similar as the former case, as both experiments rely on similar magnetic traps and probed amount of antiatoms. The necessary \dbar{} numbers would be obtained, in a 6 hours stacking run, with about 1000 \dbar / shot delivered.

\begin{table*}[t]
    \centering
    \begin{tabular}{ccc}
        \textbf{Experimental scheme} & \textbf{Impact} & \textbf{Min. est. \dbar{} flux}  \\
    \hline\hline
         \dbar{} charge-to-mass ratio in Penning trap & \dbar{} mass and binding energy, CPT/Lorentz test with \dbar{}/\nbar{} & 0.01 \dbar{} $ \mathrm{shot^{-1}} $ \\
         \dbar{} magnetic moment in Penning trap  & \dbar{} magnetic moment, CPT/Lorentz test with \dbar{}/\nbar{} & 0.01 \dbar{} $ \mathrm{shot^{-1}} $ \\
    \hline
         \dbar{} -- buffer gas mixing in a nested trap & Low-Z \dbar{}-atoms observation & 1 $ \mathrm{\bar{D} \cdot shot^{-1}} $ \\
          & Low-Z \dbar{}-atoms X-ray cascade & 10 $ \mathrm{\bar{D} \cdot shot^{-1}} $ \\         
         \dbar{} -- anion mixing with laser photodetachment & High-Z \dbar{}-atoms observation & 1 $ \mathrm{\bar{D} \cdot shot^{-1}} $ \\
         & High-Z \dbar{}-atoms X-ray cascade, low-energy \nbar{} detection & 10 $ \mathrm{\bar{D} \cdot shot^{-1}} $ \\           
    \hline
        \dbar{} -- \pos{} in nested trap & Formation of \Dbar{} & 100 $ \mathrm{\bar{d} \cdot shot^{-1}} $ \\
        \dbar{} -- \pos{} mixing in spectroscopy trap & \Dbar{} trapping and spectroscopy, CPT/Lorentz test with \Dbar{}  & 1000 $ \mathrm{\bar{d} \cdot shot^{-1}} $ \\
        \dbar{} -- \pos{} mixing in vertical trap & Gravity with \Dbar{}, constraints on long-range $ B - L $ forces & 1000 $ \mathrm{\bar{d} \cdot shot^{-1}} $ \\         
    \hline        
          \dbar{}--\dbar{} fusion in a Malmberg/Penning trap & Formation of $ \mathrm{^3 \overline{He}}$ and $ \mathrm{^3 \overline{H}}$ antinuclei & 1000 $ \mathrm{\bar{d} \cdot shot^{-1}} $ \\
          \dbar{}--\pbar{} fusion in a Malmberg/Penning trap & Formation of the $ \mathrm{^3 \overline{He}}$ antinucleus & 10000 $ \mathrm{\bar{d} \cdot shot^{-1}} $ \\          
    \end{tabular}
    \caption{Summary of the prospects enabled by experimental schemes employing cold antideuterons and estimate of the required flux}
    \label{tab:summary_impacts}
\end{table*}

\section{Synthesis of heavier antinuclei}
\label{section_heavierantinuclei}

Synthesizing heavier antinuclei by fusion reactions requires first going past the bottleneck of the

\begin{equation}
\mathrm{\bar{p} + \bar{p} \rightarrow \bar{d} + e^- + \bar{\nu}}
\end{equation}

reaction, which has a very low cross-section at \si{\kilo\electronvolt} energies (in terms of the astrophysical S-factor, $ S(0) = 4.0 \cdot 10^{-23} \, \si{\mega\electronvolt \square\femto\meter} $ \cite{Marcucci2013}). We consider here this problem solved by starting directly from a low-energy \dbar{} beam. 

We restrict our reasoning to experimental schemes technically achievable with standard technologies used by the experiments in the AD i.e., $ \approx \SI{5}{\tesla} $ radially-confining magnetic fields, cylindrical geometries with Malmberg traps featuring pulsable endcap electrodes able to reach up to $ \SI{100}{\kilo\volt} $ confining potentials \cite{Husson2021}. 

Let us consider a scenario in which both \pbar{} and \dbar{} are trapped with a minimalistic energy degrader which reduces their energy to just below the \SI{100}{\kilo\electronvolt} energy set by the ELENA ring to allow trapping multiple shots. We can consider the degraded energy distribution to be a sharp peak at $ E_0 = \SI{80}{\kilo\electronvolt} $, as shown possible by the AEgIS collaboration \footnote{unpublished work of 2022 with thin parylene-N foils}. Being no other energy loss mechanisms at play, we can assume both \pbar{} and \dbar{} to stay uncooled in the trap with a sharp energy (velocity) distribution. These are initial conditions pretty distant from a Maxwell-Boltzmann velocity distribution, usually assumed in fusion experiments. In fact, reaction rate calculations can be simplified as no averaging over reactants velocity to compute $ \langle \sigma v \rangle $ is necessary. An estimate of the reaction rate can be obtained under the approximation of head-to-head collisions between reactants with opposite directions of the momentum \footnote{this method is expected to slightly overestimate reaction rates, and could be refined by accounting for the time the reactants spent in each velocity component in the trap and the exchange between axial and radial energy components due to long-range collisions, contributing to broaden the energy distribution of the reactants.}. The cross-section are usually found in literature as $ \sigma(E_{cdm})$ where $E_{cdm} = \sqrt{s} = 2E_0$ is the center-of-mass energy. 





\subsection{Antideuteron-antiproton fusion}
The first experimental scheme we consider is antideuteron-antiproton fusion through the reaction

\begin{equation}
\mathrm{\bar{p} + \bar{d} \rightarrow \overline{^3He} + \gamma}
\end{equation}

The cross-section of the matter-counterpart of this process was recently measured at low energies by the LUNA collaboration \cite{Mossa2020, Mossa2020a}. The astrophysical S-factor at $ E_{cdm} = 2 E_0 = \SI{160}{\kilo\electronvolt} $ is $ S \approx \SI{1.3}{\electronvolt \barn}$, which corresponds to a cross-section of $ \sigma \approx \SI{1.1}{\micro\barn} $ which makes this experiment quite unrealistic. Indeed, the order of magnitude of this cross-section implies that, in order to reach a production rate of $ 1 \mathrm{\overline{^3He}} $ nucleus per week, $ 10^8 $ \pbar{} and $ 10^7 $ \dbar{} should be let interacting in a plasma of \SI{10}{\centi\meter} length and \SI{1}{\milli\meter} radius. This kind of experiment would thus necessitate the simultaneous availability of \pbar{} and \dbar{} from AD/ELENA, a \dbar{} beam intensity greater than $ 10^4 $ \dbar{} $\mathrm{shot^{-1}}$ or more and stacking for about 250 AD/ELENA shots both species at 50 \% trapping efficiency. 

\subsection{Antideuteron-antideuteron fusion}

The second experimental scheme we consider is antideuteron-antideuteron fusion in a plasma, through the set of reactions:

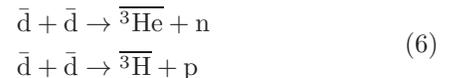
\begin{equation}
\begin{aligned}
\mathrm{\bar{d} + \bar{d}} &\rightarrow \mathrm{\overline{^3He} + n} \\
\mathrm{\bar{d} + \bar{d}} &\rightarrow \mathrm{\overline{^3H} + p} \\
\end{aligned}
\end{equation}

The combined cross-section for these two processes is much greater than the previous case, in the order of $ \sigma \approx \SI{0.1}{\barn}$ at a center of mass energy of \SI{160}{\kilo\electronvolt} \cite{Krane1987}. The $ \mathrm{\bar{d} + \bar{d} \rightarrow \overline{^4He} + \gamma}$ process instead has a negligible cross-section compared to the other two. We can estimate that, with $ 1.0 \cdot 10^6 $ \dbar{} let interacting in a plasma of \SI{10}{\centi\meter} length and \SI{1}{\milli\meter} radius, around 18 fusion events per day would be observed. With a \dbar{} beam intensity of 1000 \dbar{}$\mathrm{shot^{-1}}$ at 50 \% trapping efficiency and 6 hours of continuous accumulation of AD/ELENA shots, around $ 1.2 \cdot 10^5 $ \dbar{} would be loaded in the trap, resulting in around 1 fusion events per run (2.5 events per day). \\

\section{Feasibilty of an antideuteron beam in the AD/ELENA facility}
\label{section_dbarinad}

A first feasibility study to produce and store antideuterons was performed at the time of the Antiproton Accumulator by Johnson and Sherwood \cite{Johnson1989}. Data from several pioneering studies was used to fit a experimental antideuteron/antiproton production ratio as a function of the primary protons momentum in the laboratory frame. The fit was used to work out an expected production ratio at \SI{26}{\giga\electronvolt}/c momentum of the proton beam from PS, and found to be about $ 4 \cdot 10^{-6} $ \dbar{}/\pbar{}. An estimation of the \dbar{} momentum distribution is obtained with deuterons by both experimental data at PS \cite{Cocconi1960} and theoretical calculations \cite{Butler1963}, which indicate that the maximum of the \dbar{} momentum distribution is expected at \SI{1.7}{\giga\electronvolt}/c. 

A crude estimation of the amount of \dbar{} which can be produced from the AD target is obtained by multiplying this ratio by the current \pbar{} beam intensity in the AD, getting to $ \approx 120 $ \dbar{} per shot. Recollecting that usually 5 proton bunches from PS are used, one would conclude that about 25 \dbar{} will be produced per bunch. However, it is important to underline that this estimation disregards possible differences in the angular distribution between the two beams and acceptance considerations from the AD at a lower momentum. Furthermore, the pioneering studies on which the estimations of Johnson and Sherwood \cite{Johnson1989} are based, were obtained with lower-Z targets than what currently in use (iridium), and this can also introduce some differences. A dedicated Monte Carlo study of the \dbar{} energy and momentum distributions, accounting for the detailed geometry of the AD iridium target, is appropriate.  

From the point of view of feasibility in the current accelerator complex, it was already argued by Johnson and Sherwood that a conversion to a \dbar{} beam would be technically possible. The different \dbar{}/\pbar{} velocity would require an adaptation of the debunching cavities and the harmonic numbers in PS and AD to use multiple bunches. Stochastic and electron cooling electronics would also have to be adapted to the new revolution frequency. The real issues are linked to operation and the necessity to tune the machine with just $ 10 \div 100 $ particles. At the time of that study, no suitable technology existed; beam monitoring technology has, on the other hand, significantly evolved since then, and this aspect in particular should be re-evaluated with present-day technology.

\section{Final remarks}

We have here reviewed the physics cases opened by the development of a beam of cold antideuterons, and estimated the order of magnitude of the flux necessary to conduct several type of experiments (for a short summary of the identified physics cases and the relative necessary fluxes of \dbar{} needed, see Table \ref{tab:summary_impacts}). Antideuteron beams can play a fundamental role in opening up a new field of experiments involving antineutrons. In the long term, if sufficient fluxes can be reached, trapped \dbar{}s may play the role of initiators of anti-nucleosynthesis i.e., allowing to synthesize heavier anti-nuclei by fusion proceses in low-energy experiments.

The estimated number of \dbar{}s produced by the AD target is several orders of magnitude higher than the needs of some experimental schemes listed in this review. On the other hand, the feasibility of collecting and cooling these antideuterons in the current AD/ELENA facility should be further investigated in dedicated studies focusing on the technical aspects of such a challenging development. 

\section*{Acknowledgments}

The author is grateful to D. Gamba and M. Doser for the very insightful discussions. This work was financially supported by Istituto Nazionale di Fisica Nucleare.

\bibliographystyle{apsrev}
\bibliography{library_antideuteron}

\begin{thebibliography}{39}
\expandafter\ifx\csname natexlab\endcsname\relax\def\natexlab#1{#1}\fi
\expandafter\ifx\csname bibnamefont\endcsname\relax
  \def\bibnamefont#1{#1}\fi
\expandafter\ifx\csname bibfnamefont\endcsname\relax
  \def\bibfnamefont#1{#1}\fi
\expandafter\ifx\csname citenamefont\endcsname\relax
  \def\citenamefont#1{#1}\fi
\expandafter\ifx\csname url\endcsname\relax
  \def\url#1{\texttt{#1}}\fi
\expandafter\ifx\csname urlprefix\endcsname\relax\def\urlprefix{URL }\fi
\providecommand{\bibinfo}[2]{#2}
\providecommand{\eprint}[2][]{\url{#2}}

\bibitem[{\citenamefont{Acharya et~al.}(2022)\citenamefont{Acharya, Adamová,
  Adler, Adolfsson, Aglieri~Rinella, Agnello, Agrawal, Ahammed, Ahmad, Ahn
  et~al.}}]{ALICE2022Antinuclei}
\bibinfo{author}{\bibfnamefont{S.}~\bibnamefont{Acharya}},
  \bibinfo{author}{\bibfnamefont{D.}~\bibnamefont{Adamová}},
  \bibinfo{author}{\bibfnamefont{A.}~\bibnamefont{Adler}},
  \bibinfo{author}{\bibfnamefont{J.}~\bibnamefont{Adolfsson}},
  \bibinfo{author}{\bibfnamefont{G.}~\bibnamefont{Aglieri~Rinella}},
  \bibinfo{author}{\bibfnamefont{M.}~\bibnamefont{Agnello}},
  \bibinfo{author}{\bibfnamefont{N.}~\bibnamefont{Agrawal}},
  \bibinfo{author}{\bibfnamefont{Z.}~\bibnamefont{Ahammed}},
  \bibinfo{author}{\bibfnamefont{S.}~\bibnamefont{Ahmad}},
  \bibinfo{author}{\bibfnamefont{S.~U.} \bibnamefont{Ahn}},
  \bibnamefont{et~al.}, \bibinfo{journal}{Journal of High Energy Physics}
  \textbf{\bibinfo{volume}{2022}} (\bibinfo{year}{2022}), ISSN
  \bibinfo{issn}{1029-8479}.

\bibitem[{\citenamefont{Massam et~al.}(1965)\citenamefont{Massam, Muller,
  Righini, Schneegans, and Zichichi}}]{Zichichi1965}
\bibinfo{author}{\bibfnamefont{T.}~\bibnamefont{Massam}},
  \bibinfo{author}{\bibfnamefont{T.}~\bibnamefont{Muller}},
  \bibinfo{author}{\bibfnamefont{B.}~\bibnamefont{Righini}},
  \bibinfo{author}{\bibfnamefont{M.}~\bibnamefont{Schneegans}},
  \bibnamefont{and} \bibinfo{author}{\bibfnamefont{A.}~\bibnamefont{Zichichi}},
  \bibinfo{journal}{Il Nuovo Cimento A} \textbf{\bibinfo{volume}{63}},
  \bibinfo{pages}{10} (\bibinfo{year}{1965}), ISSN \bibinfo{issn}{1826-9869}.

\bibitem[{\citenamefont{Dorfan et~al.}(1965)\citenamefont{Dorfan, Eades,
  Lederman, Lee, and Ting}}]{Dorfan1965}
\bibinfo{author}{\bibfnamefont{D.~E.} \bibnamefont{Dorfan}},
  \bibinfo{author}{\bibfnamefont{J.}~\bibnamefont{Eades}},
  \bibinfo{author}{\bibfnamefont{L.~M.} \bibnamefont{Lederman}},
  \bibinfo{author}{\bibfnamefont{W.}~\bibnamefont{Lee}}, \bibnamefont{and}
  \bibinfo{author}{\bibfnamefont{C.~C.} \bibnamefont{Ting}},
  \bibinfo{journal}{Physical Review Letters} \textbf{\bibinfo{volume}{14}},
  \bibinfo{pages}{1003} (\bibinfo{year}{1965}), ISSN \bibinfo{issn}{0031-9007}.

\bibitem[{\citenamefont{Möhl et~al.}(1982)\citenamefont{Möhl, Kilian,
  Pilkuhn, and Poth}}]{Moehl1982}
\bibinfo{author}{\bibfnamefont{D.}~\bibnamefont{Möhl}},
  \bibinfo{author}{\bibfnamefont{K.}~\bibnamefont{Kilian}},
  \bibinfo{author}{\bibfnamefont{H.}~\bibnamefont{Pilkuhn}}, \bibnamefont{and}
  \bibinfo{author}{\bibfnamefont{H.}~\bibnamefont{Poth}},
  \bibinfo{journal}{Nuclear Instruments and Methods in Physics Research}
  \textbf{\bibinfo{volume}{202}}, \bibinfo{pages}{427} (\bibinfo{year}{1982}),
  ISSN \bibinfo{issn}{0167-5087}.

\bibitem[{\citenamefont{Koch et~al.}(1984)\citenamefont{Koch, Kilian, Möhl,
  Pilkuhn, and Poth}}]{Koch1984}
\bibinfo{author}{\bibfnamefont{H.}~\bibnamefont{Koch}},
  \bibinfo{author}{\bibfnamefont{K.}~\bibnamefont{Kilian}},
  \bibinfo{author}{\bibfnamefont{D.}~\bibnamefont{Möhl}},
  \bibinfo{author}{\bibfnamefont{H.}~\bibnamefont{Pilkuhn}}, \bibnamefont{and}
  \bibinfo{author}{\bibfnamefont{H.}~\bibnamefont{Poth}}, in
  \emph{\bibinfo{booktitle}{Physics at LEAR with Low-Energy Cooled
  Antiprotons}}, edited by
  \bibinfo{editor}{\bibfnamefont{U.}~\bibnamefont{Gastaldi}}
  (\bibinfo{publisher}{Springer}, \bibinfo{year}{1984}).

\bibitem[{\citenamefont{Johnson and Sherwood}(1989)}]{Johnson1989}
\bibinfo{author}{\bibfnamefont{C.~D.} \bibnamefont{Johnson}} \bibnamefont{and}
  \bibinfo{author}{\bibfnamefont{T.~R.} \bibnamefont{Sherwood}},
  \bibinfo{journal}{Hyperfine Interactions} \textbf{\bibinfo{volume}{44}},
  \bibinfo{pages}{65} (\bibinfo{year}{1989}), ISSN \bibinfo{issn}{1572-9540}.

\bibitem[{\citenamefont{Koch}(1989)}]{Koch1989}
\bibinfo{author}{\bibfnamefont{H.}~\bibnamefont{Koch}},
  \bibinfo{journal}{Hyperfine Interactions} \textbf{\bibinfo{volume}{44}},
  \bibinfo{pages}{59} (\bibinfo{year}{1989}), ISSN \bibinfo{issn}{1572-9540}.

\bibitem[{Note1()}]{Note1}
Note1, \bibinfo{note}{no explicit reference to measurements done at colliders
  found yet.}

\bibitem[{\citenamefont{Rau et~al.}(2020)\citenamefont{Rau, Heiße,
  Köhler-Langes, Sasidharan, Haas, Renisch, Düllmann, Quint, Sturm, and
  Blaum}}]{Rau2020}
\bibinfo{author}{\bibfnamefont{S.}~\bibnamefont{Rau}},
  \bibinfo{author}{\bibfnamefont{F.}~\bibnamefont{Heiße}},
  \bibinfo{author}{\bibfnamefont{F.}~\bibnamefont{Köhler-Langes}},
  \bibinfo{author}{\bibfnamefont{S.}~\bibnamefont{Sasidharan}},
  \bibinfo{author}{\bibfnamefont{R.}~\bibnamefont{Haas}},
  \bibinfo{author}{\bibfnamefont{D.}~\bibnamefont{Renisch}},
  \bibinfo{author}{\bibfnamefont{C.~E.} \bibnamefont{Düllmann}},
  \bibinfo{author}{\bibfnamefont{W.}~\bibnamefont{Quint}},
  \bibinfo{author}{\bibfnamefont{S.}~\bibnamefont{Sturm}}, \bibnamefont{and}
  \bibinfo{author}{\bibfnamefont{K.}~\bibnamefont{Blaum}},
  \bibinfo{journal}{Nature} \textbf{\bibinfo{volume}{585}}, \bibinfo{pages}{43}
  (\bibinfo{year}{2020}), ISSN \bibinfo{issn}{1476-4687}.

\bibitem[{\citenamefont{Borchert et~al.}(2022)\citenamefont{Borchert, Devlin,
  Erlewein, Fleck, Harrington, Higuchi, Latacz, Voelksen, Wursten, Abbass
  et~al.}}]{BASE2022}
\bibinfo{author}{\bibfnamefont{M.~J.} \bibnamefont{Borchert}},
  \bibinfo{author}{\bibfnamefont{J.~A.} \bibnamefont{Devlin}},
  \bibinfo{author}{\bibfnamefont{S.~R.} \bibnamefont{Erlewein}},
  \bibinfo{author}{\bibfnamefont{M.}~\bibnamefont{Fleck}},
  \bibinfo{author}{\bibfnamefont{J.~A.} \bibnamefont{Harrington}},
  \bibinfo{author}{\bibfnamefont{T.}~\bibnamefont{Higuchi}},
  \bibinfo{author}{\bibfnamefont{B.~M.} \bibnamefont{Latacz}},
  \bibinfo{author}{\bibfnamefont{F.}~\bibnamefont{Voelksen}},
  \bibinfo{author}{\bibfnamefont{E.~J.} \bibnamefont{Wursten}},
  \bibinfo{author}{\bibfnamefont{F.}~\bibnamefont{Abbass}},
  \bibnamefont{et~al.}, \bibinfo{journal}{Nature}
  \textbf{\bibinfo{volume}{601}}, \bibinfo{pages}{53} (\bibinfo{year}{2022}),
  ISSN \bibinfo{issn}{1476-4687}.

\bibitem[{\citenamefont{Latasz}()}]{BASEPrivateConv}
\bibinfo{author}{\bibfnamefont{B.}~\bibnamefont{Latasz}}.

\bibitem[{\citenamefont{Krane}(1987)}]{Krane1987}
\bibinfo{author}{\bibfnamefont{K.~S.} \bibnamefont{Krane}},
  \emph{\bibinfo{title}{Introductory nuclear physics}}
  (\bibinfo{publisher}{Wiley}, \bibinfo{year}{1987}), ISBN
  \bibinfo{isbn}{047180553X}.

\bibitem[{\citenamefont{Anastassopoulos et~al.}(2008)}]{Anastassopoulos2008}
\bibinfo{author}{\bibfnamefont{D.}~\bibnamefont{Anastassopoulos}}
  \bibnamefont{et~al.}, \bibinfo{type}{Tech. Rep.}, \bibinfo{institution}{BNL}
  (\bibinfo{year}{2008}).

\bibitem[{\citenamefont{Smorra et~al.}(2017)\citenamefont{Smorra, Sellner,
  Borchert, Harrington, Higuchi, Nagahama, Tanaka, Mooser, Schneider, Bohman
  et~al.}}]{Smorra2017}
\bibinfo{author}{\bibfnamefont{C.}~\bibnamefont{Smorra}},
  \bibinfo{author}{\bibfnamefont{S.}~\bibnamefont{Sellner}},
  \bibinfo{author}{\bibfnamefont{M.~J.} \bibnamefont{Borchert}},
  \bibinfo{author}{\bibfnamefont{J.~A.} \bibnamefont{Harrington}},
  \bibinfo{author}{\bibfnamefont{T.}~\bibnamefont{Higuchi}},
  \bibinfo{author}{\bibfnamefont{H.}~\bibnamefont{Nagahama}},
  \bibinfo{author}{\bibfnamefont{T.}~\bibnamefont{Tanaka}},
  \bibinfo{author}{\bibfnamefont{A.}~\bibnamefont{Mooser}},
  \bibinfo{author}{\bibfnamefont{G.}~\bibnamefont{Schneider}},
  \bibinfo{author}{\bibfnamefont{M.}~\bibnamefont{Bohman}},
  \bibnamefont{et~al.}, \bibinfo{journal}{Nature}
  \textbf{\bibinfo{volume}{550}}, \bibinfo{pages}{371} (\bibinfo{year}{2017}),
  ISSN \bibinfo{issn}{1476-4687}.

\bibitem[{\citenamefont{Bressani}(2003)}]{Bressani2003}
\bibinfo{author}{\bibfnamefont{T.}~\bibnamefont{Bressani}},
  \bibinfo{journal}{Physics Reports} \textbf{\bibinfo{volume}{383}},
  \bibinfo{pages}{213} (\bibinfo{year}{2003}), ISSN \bibinfo{issn}{0370-1573}.

\bibitem[{\citenamefont{Hannaske et~al.}(2016)\citenamefont{Hannaske, Bemmerer,
  Beyer, Birgersson, Ferrari, Grosse, Junghans, Kempe, Kögler, Kosev
  et~al.}}]{Hannaske2016}
\bibinfo{author}{\bibfnamefont{R.}~\bibnamefont{Hannaske}},
  \bibinfo{author}{\bibfnamefont{D.}~\bibnamefont{Bemmerer}},
  \bibinfo{author}{\bibfnamefont{R.}~\bibnamefont{Beyer}},
  \bibinfo{author}{\bibfnamefont{E.}~\bibnamefont{Birgersson}},
  \bibinfo{author}{\bibfnamefont{A.}~\bibnamefont{Ferrari}},
  \bibinfo{author}{\bibfnamefont{E.}~\bibnamefont{Grosse}},
  \bibinfo{author}{\bibfnamefont{A.~R.} \bibnamefont{Junghans}},
  \bibinfo{author}{\bibfnamefont{M.}~\bibnamefont{Kempe}},
  \bibinfo{author}{\bibfnamefont{T.}~\bibnamefont{Kögler}},
  \bibinfo{author}{\bibfnamefont{K.}~\bibnamefont{Kosev}},
  \bibnamefont{et~al.}, \bibinfo{journal}{Journal of Physics: Conference
  Series} \textbf{\bibinfo{volume}{665}}, \bibinfo{pages}{012003}
  (\bibinfo{year}{2016}), ISSN \bibinfo{issn}{1742-6596}.

\bibitem[{\citenamefont{Baur and Trautmann}(1972)}]{Baur1972}
\bibinfo{author}{\bibfnamefont{G.}~\bibnamefont{Baur}} \bibnamefont{and}
  \bibinfo{author}{\bibfnamefont{D.}~\bibnamefont{Trautmann}},
  \bibinfo{journal}{Physics Letters B} \textbf{\bibinfo{volume}{42}},
  \bibinfo{pages}{31} (\bibinfo{year}{1972}), ISSN \bibinfo{issn}{0370-2693}.

\bibitem[{\citenamefont{Ericson and Osland}(1975)}]{Ericson1975}
\bibinfo{author}{\bibfnamefont{T.}~\bibnamefont{Ericson}} \bibnamefont{and}
  \bibinfo{author}{\bibfnamefont{P.}~\bibnamefont{Osland}},
  \bibinfo{journal}{Nuclear Physics A} \textbf{\bibinfo{volume}{249}},
  \bibinfo{pages}{445} (\bibinfo{year}{1975}), ISSN \bibinfo{issn}{0375-9474}.

\bibitem[{Note2()}]{Note2}
Note2, \bibinfo{note}{unpublished}.

\bibitem[{\citenamefont{Welsh}(1989)}]{Welsh1989}
\bibinfo{author}{\bibfnamefont{R.~E.} \bibnamefont{Welsh}},
  \bibinfo{journal}{Nuclear Physics B - Proceedings Supplements}
  \textbf{\bibinfo{volume}{8}}, \bibinfo{pages}{90} (\bibinfo{year}{1989}),
  ISSN \bibinfo{issn}{0920-5632}.

\bibitem[{\citenamefont{Gastaldi}(1978)}]{Gastaldi1978}
\bibinfo{author}{\bibfnamefont{U.}~\bibnamefont{Gastaldi}},
  \bibinfo{journal}{Nuclear Instruments and Methods}
  \textbf{\bibinfo{volume}{157}}, \bibinfo{pages}{441} (\bibinfo{year}{1978}),
  ISSN \bibinfo{issn}{0029-554X}.

\bibitem[{\citenamefont{Baur}(1976)}]{Baur1976}
\bibinfo{author}{\bibfnamefont{G.}~\bibnamefont{Baur}},
  \bibinfo{journal}{Physics Letters B} \textbf{\bibinfo{volume}{60}},
  \bibinfo{pages}{137} (\bibinfo{year}{1976}), ISSN \bibinfo{issn}{0370-2693}.

\bibitem[{\citenamefont{Middleton}(1983)}]{Middleton1983}
\bibinfo{author}{\bibfnamefont{R.}~\bibnamefont{Middleton}},
  \bibinfo{journal}{Nuclear Instruments and Methods in Physics Research}
  \textbf{\bibinfo{volume}{214}}, \bibinfo{pages}{139} (\bibinfo{year}{1983}),
  ISSN \bibinfo{issn}{0167-5087}.

\bibitem[{Note3()}]{Note3}
Note3, \bibinfo{note}{yet to be calculated}.

\bibitem[{\citenamefont{Amoretti et~al.}(2002)\citenamefont{Amoretti, Amsler,
  Bonomi, Bouchta, Bowe, Carraro, Cesar, Charlton, Collier, Doser
  et~al.}}]{Amoretti2002}
\bibinfo{author}{\bibfnamefont{M.}~\bibnamefont{Amoretti}},
  \bibinfo{author}{\bibfnamefont{C.}~\bibnamefont{Amsler}},
  \bibinfo{author}{\bibfnamefont{G.}~\bibnamefont{Bonomi}},
  \bibinfo{author}{\bibfnamefont{A.}~\bibnamefont{Bouchta}},
  \bibinfo{author}{\bibfnamefont{P.}~\bibnamefont{Bowe}},
  \bibinfo{author}{\bibfnamefont{C.}~\bibnamefont{Carraro}},
  \bibinfo{author}{\bibfnamefont{C.~L.} \bibnamefont{Cesar}},
  \bibinfo{author}{\bibfnamefont{M.}~\bibnamefont{Charlton}},
  \bibinfo{author}{\bibfnamefont{M.~J.~T.} \bibnamefont{Collier}},
  \bibinfo{author}{\bibfnamefont{M.}~\bibnamefont{Doser}},
  \bibnamefont{et~al.}, \bibinfo{journal}{Nature}
  \textbf{\bibinfo{volume}{419}}, \bibinfo{pages}{456} (\bibinfo{year}{2002}),
  ISSN \bibinfo{issn}{1476-4687}.

\bibitem[{\citenamefont{Wineland and Ramsey}(1972)}]{Wineland1972}
\bibinfo{author}{\bibfnamefont{D.~J.} \bibnamefont{Wineland}} \bibnamefont{and}
  \bibinfo{author}{\bibfnamefont{N.~F.} \bibnamefont{Ramsey}},
  \bibinfo{journal}{Physical Review A} \textbf{\bibinfo{volume}{5}},
  \bibinfo{pages}{821} (\bibinfo{year}{1972}), ISSN \bibinfo{issn}{0556-2791}.

\bibitem[{\citenamefont{Andresen et~al.}(2010)\citenamefont{Andresen,
  Ashkezari, Baquero-Ruiz, Bertsche, Bowe, Butler, Cesar, Chapman, Charlton,
  Deller et~al.}}]{Andresen2010}
\bibinfo{author}{\bibfnamefont{G.~B.} \bibnamefont{Andresen}},
  \bibinfo{author}{\bibfnamefont{M.~D.} \bibnamefont{Ashkezari}},
  \bibinfo{author}{\bibfnamefont{M.}~\bibnamefont{Baquero-Ruiz}},
  \bibinfo{author}{\bibfnamefont{W.}~\bibnamefont{Bertsche}},
  \bibinfo{author}{\bibfnamefont{P.~D.} \bibnamefont{Bowe}},
  \bibinfo{author}{\bibfnamefont{E.}~\bibnamefont{Butler}},
  \bibinfo{author}{\bibfnamefont{C.~L.} \bibnamefont{Cesar}},
  \bibinfo{author}{\bibfnamefont{S.}~\bibnamefont{Chapman}},
  \bibinfo{author}{\bibfnamefont{M.}~\bibnamefont{Charlton}},
  \bibinfo{author}{\bibfnamefont{A.}~\bibnamefont{Deller}},
  \bibnamefont{et~al.}, \bibinfo{journal}{Nature}
  \textbf{\bibinfo{volume}{468}}, \bibinfo{pages}{673} (\bibinfo{year}{2010}),
  ISSN \bibinfo{issn}{1476-4687}.

\bibitem[{\citenamefont{Ahmadi et~al.}(2017)\citenamefont{Ahmadi, Alves, Baker,
  Bertsche, Butler, Capra, Carruth, Cesar, Charlton, Cohen
  et~al.}}]{Ahmadi2017}
\bibinfo{author}{\bibfnamefont{M.}~\bibnamefont{Ahmadi}},
  \bibinfo{author}{\bibfnamefont{B.~X.~R.} \bibnamefont{Alves}},
  \bibinfo{author}{\bibfnamefont{C.~J.} \bibnamefont{Baker}},
  \bibinfo{author}{\bibfnamefont{W.}~\bibnamefont{Bertsche}},
  \bibinfo{author}{\bibfnamefont{E.}~\bibnamefont{Butler}},
  \bibinfo{author}{\bibfnamefont{A.}~\bibnamefont{Capra}},
  \bibinfo{author}{\bibfnamefont{C.}~\bibnamefont{Carruth}},
  \bibinfo{author}{\bibfnamefont{C.~L.} \bibnamefont{Cesar}},
  \bibinfo{author}{\bibfnamefont{M.}~\bibnamefont{Charlton}},
  \bibinfo{author}{\bibfnamefont{S.}~\bibnamefont{Cohen}},
  \bibnamefont{et~al.}, \bibinfo{journal}{Nature Communications}
  \textbf{\bibinfo{volume}{8}} (\bibinfo{year}{2017}), ISSN
  \bibinfo{issn}{2041-1723}.

\bibitem[{\citenamefont{Charlton et~al.}(2020)\citenamefont{Charlton, Eriksson,
  and Shore}}]{Charlton2020}
\bibinfo{author}{\bibfnamefont{M.}~\bibnamefont{Charlton}},
  \bibinfo{author}{\bibfnamefont{S.}~\bibnamefont{Eriksson}}, \bibnamefont{and}
  \bibinfo{author}{\bibfnamefont{G.~M.} \bibnamefont{Shore}},
  \emph{\bibinfo{title}{Antihydrogen and Fundamental Physics : Testing
  Fundamental Physics}} (\bibinfo{publisher}{Springer}, \bibinfo{year}{2020}),
  ISBN \bibinfo{isbn}{9783030517120}.

\bibitem[{\citenamefont{Heeck}(2014)}]{Heeck2014}
\bibinfo{author}{\bibfnamefont{J.}~\bibnamefont{Heeck}},
  \bibinfo{journal}{Physics Letters B} \textbf{\bibinfo{volume}{739}},
  \bibinfo{pages}{256} (\bibinfo{year}{2014}), ISSN \bibinfo{issn}{0370-2693}.

\bibitem[{\citenamefont{Anderson et~al.}(2023)\citenamefont{Anderson, Baker,
  Bertsche, Bhatt, Bonomi, Capra, Carli, Cesar, Charlton, Christensen
  et~al.}}]{Anderson2023}
\bibinfo{author}{\bibfnamefont{E.~K.} \bibnamefont{Anderson}},
  \bibinfo{author}{\bibfnamefont{C.~J.} \bibnamefont{Baker}},
  \bibinfo{author}{\bibfnamefont{W.}~\bibnamefont{Bertsche}},
  \bibinfo{author}{\bibfnamefont{N.~M.} \bibnamefont{Bhatt}},
  \bibinfo{author}{\bibfnamefont{G.}~\bibnamefont{Bonomi}},
  \bibinfo{author}{\bibfnamefont{A.}~\bibnamefont{Capra}},
  \bibinfo{author}{\bibfnamefont{I.}~\bibnamefont{Carli}},
  \bibinfo{author}{\bibfnamefont{C.~L.} \bibnamefont{Cesar}},
  \bibinfo{author}{\bibfnamefont{M.}~\bibnamefont{Charlton}},
  \bibinfo{author}{\bibfnamefont{A.}~\bibnamefont{Christensen}},
  \bibnamefont{et~al.}, \bibinfo{journal}{Nature}
  \textbf{\bibinfo{volume}{621}}, \bibinfo{pages}{716â€“722}
  (\bibinfo{year}{2023}), ISSN \bibinfo{issn}{1476-4687}.

\bibitem[{\citenamefont{Marcucci et~al.}(2013)\citenamefont{Marcucci,
  Schiavilla, and Viviani}}]{Marcucci2013}
\bibinfo{author}{\bibfnamefont{L.~E.} \bibnamefont{Marcucci}},
  \bibinfo{author}{\bibfnamefont{R.}~\bibnamefont{Schiavilla}},
  \bibnamefont{and} \bibinfo{author}{\bibfnamefont{M.}~\bibnamefont{Viviani}},
  \bibinfo{journal}{Physical Review Letters} \textbf{\bibinfo{volume}{110}},
  \bibinfo{pages}{192503} (\bibinfo{year}{2013}), ISSN
  \bibinfo{issn}{1079-7114}.

\bibitem[{\citenamefont{Husson et~al.}(2021)\citenamefont{Husson, Kim, Welker,
  Charlton, Choi, Chung, Cladé, Comini, Crépin, Crivelli
  et~al.}}]{Husson2021}
\bibinfo{author}{\bibfnamefont{A.}~\bibnamefont{Husson}},
  \bibinfo{author}{\bibfnamefont{B.}~\bibnamefont{Kim}},
  \bibinfo{author}{\bibfnamefont{A.}~\bibnamefont{Welker}},
  \bibinfo{author}{\bibfnamefont{M.}~\bibnamefont{Charlton}},
  \bibinfo{author}{\bibfnamefont{J.}~\bibnamefont{Choi}},
  \bibinfo{author}{\bibfnamefont{M.}~\bibnamefont{Chung}},
  \bibinfo{author}{\bibfnamefont{P.}~\bibnamefont{Cladé}},
  \bibinfo{author}{\bibfnamefont{P.}~\bibnamefont{Comini}},
  \bibinfo{author}{\bibfnamefont{P.-P.} \bibnamefont{Crépin}},
  \bibinfo{author}{\bibfnamefont{P.}~\bibnamefont{Crivelli}},
  \bibnamefont{et~al.}, \bibinfo{journal}{Nuclear Instruments and Methods in
  Physics Research Section A: Accelerators, Spectrometers, Detectors and
  Associated Equipment} \textbf{\bibinfo{volume}{1002}},
  \bibinfo{pages}{165245} (\bibinfo{year}{2021}), ISSN
  \bibinfo{issn}{0168-9002}.

\bibitem[{Note4()}]{Note4}
Note4, \bibinfo{note}{unpublished work of 2022 with thin parylene-N foils}.

\bibitem[{Note5()}]{Note5}
Note5, \bibinfo{note}{this method is expected to slightly overestimate reaction
  rates, and could be refined by accounting for the time the reactants spent in
  each velocity component in the trap and the exchange between axial and radial
  energy components due to long-range collisions, contributing to broaden the
  energy distribution of the reactants.}

\bibitem[{\citenamefont{Mossa et~al.}(2020{\natexlab{a}})\citenamefont{Mossa,
  Stöckel, Cavanna, Ferraro, Aliotta, Barile, Bemmerer, Best, Boeltzig,
  Broggini et~al.}}]{Mossa2020}
\bibinfo{author}{\bibfnamefont{V.}~\bibnamefont{Mossa}},
  \bibinfo{author}{\bibfnamefont{K.}~\bibnamefont{Stöckel}},
  \bibinfo{author}{\bibfnamefont{F.}~\bibnamefont{Cavanna}},
  \bibinfo{author}{\bibfnamefont{F.}~\bibnamefont{Ferraro}},
  \bibinfo{author}{\bibfnamefont{M.}~\bibnamefont{Aliotta}},
  \bibinfo{author}{\bibfnamefont{F.}~\bibnamefont{Barile}},
  \bibinfo{author}{\bibfnamefont{D.}~\bibnamefont{Bemmerer}},
  \bibinfo{author}{\bibfnamefont{A.}~\bibnamefont{Best}},
  \bibinfo{author}{\bibfnamefont{A.}~\bibnamefont{Boeltzig}},
  \bibinfo{author}{\bibfnamefont{C.}~\bibnamefont{Broggini}},
  \bibnamefont{et~al.}, \bibinfo{journal}{Nature}
  \textbf{\bibinfo{volume}{587}}, \bibinfo{pages}{210}
  (\bibinfo{year}{2020}{\natexlab{a}}), ISSN \bibinfo{issn}{1476-4687}.

\bibitem[{\citenamefont{Mossa et~al.}(2020{\natexlab{b}})\citenamefont{Mossa,
  Stöckel, Cavanna, Ferraro, Aliotta, Barile, Bemmerer, Best, Boeltzig,
  Broggini et~al.}}]{Mossa2020a}
\bibinfo{author}{\bibfnamefont{V.}~\bibnamefont{Mossa}},
  \bibinfo{author}{\bibfnamefont{K.}~\bibnamefont{Stöckel}},
  \bibinfo{author}{\bibfnamefont{F.}~\bibnamefont{Cavanna}},
  \bibinfo{author}{\bibfnamefont{F.}~\bibnamefont{Ferraro}},
  \bibinfo{author}{\bibfnamefont{M.}~\bibnamefont{Aliotta}},
  \bibinfo{author}{\bibfnamefont{F.}~\bibnamefont{Barile}},
  \bibinfo{author}{\bibfnamefont{D.}~\bibnamefont{Bemmerer}},
  \bibinfo{author}{\bibfnamefont{A.}~\bibnamefont{Best}},
  \bibinfo{author}{\bibfnamefont{A.}~\bibnamefont{Boeltzig}},
  \bibinfo{author}{\bibfnamefont{C.}~\bibnamefont{Broggini}},
  \bibnamefont{et~al.}, \bibinfo{journal}{The European Physical Journal A}
  \textbf{\bibinfo{volume}{56}} (\bibinfo{year}{2020}{\natexlab{b}}), ISSN
  \bibinfo{issn}{1434-601X}.

\bibitem[{\citenamefont{Cocconi et~al.}(1960)\citenamefont{Cocconi, Fazzini,
  Fidecaro, Legros, Lipman, and Merrison}}]{Cocconi1960}
\bibinfo{author}{\bibfnamefont{V.~T.} \bibnamefont{Cocconi}},
  \bibinfo{author}{\bibfnamefont{T.}~\bibnamefont{Fazzini}},
  \bibinfo{author}{\bibfnamefont{G.}~\bibnamefont{Fidecaro}},
  \bibinfo{author}{\bibfnamefont{M.}~\bibnamefont{Legros}},
  \bibinfo{author}{\bibfnamefont{N.~H.} \bibnamefont{Lipman}},
  \bibnamefont{and} \bibinfo{author}{\bibfnamefont{A.~W.}
  \bibnamefont{Merrison}}, \bibinfo{journal}{Physical Review Letters}
  \textbf{\bibinfo{volume}{5}}, \bibinfo{pages}{19} (\bibinfo{year}{1960}),
  ISSN \bibinfo{issn}{0031-9007}.

\bibitem[{\citenamefont{Butler and Pearson}(1963)}]{Butler1963}
\bibinfo{author}{\bibfnamefont{S.~T.} \bibnamefont{Butler}} \bibnamefont{and}
  \bibinfo{author}{\bibfnamefont{C.~A.} \bibnamefont{Pearson}},
  \bibinfo{journal}{Physical Review} \textbf{\bibinfo{volume}{129}},
  \bibinfo{pages}{836} (\bibinfo{year}{1963}), ISSN \bibinfo{issn}{0031-899X}.

\end{thebibliography}

\end{document}